\documentclass[12pt]{iopart}
\usepackage{graphicx}
\usepackage{iopams}
\usepackage{psfrag} 
\usepackage{fancyhdr}
\begin{document}
\date{\today}
\title{Kinetics of charge inversion}
\author{\bf Yan Levin, Jeferson J. Arenzon} 
\address{\it Instituto de F\'{\i}sica, Universidade Federal
do Rio Grande do Sul\\ Caixa Postal 15051, CEP 91501-970, 
Porto Alegre, RS, Brazil\\ 
{\small levin@if.ufrgs.br}}

\begin{abstract}

Colloidal suspensions and polyelectrolyte solutions containing
multivalent counterions can exhibit some very counter-intuitive
behavior usually associated with the low temperature physics. 
There are two particularly striking phenomena resulting from strong
electrostatic correlations. One is the like-charge attraction and the
second is the polyion overcharging.  
In this contribution we will concentrate on the problem of overcharging.
In particular we will explore the kinetic limitation to colloidal
charge inversion in suspensions containing multivalent counterions.

\end{abstract}
\maketitle
\bigskip
\pagestyle{fancy}
\lhead{\thepage}
\rhead{Kinetics of charge inversion}
\cfoot{}

\section{Introduction}

Colloidal suspensions and polyelectrolyte solutions containing
multivalent counterions can exhibit some very curious  electrostatic
behavior~\cite{Le02}.  It is found that under some 
circumstances two like-charged
polyions inside suspension can actually attract one 
another~\cite{Pa80,GuJoWe84,KjMa86,StRo90,CrGr94,RoBl96,HaLi97,LeArSt99,GrMaBr97,KoLe99,AlAmLo98,ArStLe99,GoHa99,LaPiLeFe00,DiCaLe01}. 
The counterion mediated attraction is responsible for the $DNA$ compaction
inside the bacteriophages, viruses that infect
bacteria~\cite{Bl91,Bl97}, and for the organization of 
eukaryotic cytoskeleton~\cite{TaJa96}.
Another ``strange'' electrostatic behavior which can occur in
suspensions containing multivalent counterions is the reversal of the
electrophoretic 
mobility~\cite{Le02,LoSaHe83,GoLoHe85,Sh99a,NgGrSh00b,MeHoKr01,GrNgSh02}.  
The first thing that is learned in a course on electrostatics
is that the force produced by the electric field on a charged particle is
\begin{equation}
\label{1}
\mathbf{F}=Q \mathbf{E} \;.  
\end{equation}
Thus, a positively charged particle, $Q>0$, is expected to 
move in the direction
of the applied field while a negatively charged particle, $Q<0$,
will move in the direction opposite to the field. This simple
picture, however, breaks down inside colloidal
suspensions of low dielectric solvent or even in 
aqueous suspensions
containing multivalent counterions.  The reason for the  violation
of the ``simple'' physics learned in high-school are the strong
electrostatic many-body interactions between the colloidal particles
and the counterions.  The reversal of electrophoretic mobility
can be understood as a combination of two electrostatically driven
mechanisms.  Strong electrostatic interaction between
colloids and counterions  leads
to formation of polyion-counterion 
complexes~\cite{AlChGr84,LeBaTa98,DiBaLe01}. The 
existence of counterion condensation has 
been known for over thirty years~\cite{Ma69,Ma78,Oo71},  
the general phenomenon is, however, much older than this
and can be traced to the pioneering work of Bjerrum on
ionic association inside electrolyte solutions 
almost $80$ years ago~\cite{Bj26}.
In aqueous suspensions with only monovalent counterions, the net charge
of complexes is of the same sign
as the bare charge of polyions.

If the solvent is water and the counterions are monovalent, 
the electrostatic
interactions between the condensed counterions can be neglected~\cite{Le02},
and the simplest Poisson-Boltzmann  theory is  sufficient
to describe the polyion-counterion complexation~\cite{AlChGr84,TrBoAu02}. 
In aqueous suspensions containing multivalent counterions or
in  suspension of low
dielectric solvents, the electrostatic energy between the condensed
counterions is significantly larger than the thermal energy and the 
electrostatic correlations between the condensed counterions can no
longer be neglected. These electrostatic correlations 
can lead to colloidal overcharging i.e. the  net charge of
the complex is of opposite sign to the charge of the bare polyion.  
The overcharged colloid will then move in the ``wrong'' direction
with respect to the applied electric field~\cite{Le02,GrNgSh02}.

\section{Overcharging}

To understand the phenomenon of overcharging we shall start by studying
a very simple model.  Consider a sphere of radius $a$ and 
fixed charge $-Zq$ distributed uniformly
over its surface. We would like to know how 
many point-like $\alpha$-valent counterions, each of charge $\alpha q$, 
should be placed
on top of this sphere in order to minimize the total electrostatic
free energy~\cite{Th04,Le02,Sh99a,MeHoKr01}. 
When we say ``counterions'' we have in mind both simple
multivalent ions such as $Ca^{++}$, as well as more complicated
micelle-like aggregates with $\alpha$ significantly higher than one. 

The free energy of a complex can be written as  
\begin{equation}
\label{2} 
E_n=\frac{Z^2 q^2}{2 \epsilon a}-
\frac{Z \alpha n q^2}{\epsilon a}+F^{\alpha \alpha}_n \;.
\end{equation}
The first term is the self energy of the charged sphere,  the second
term is the electrostatic energy of interaction between the sphere
and $n$ condensed $\alpha$-ions, and  the last term is the
electrostatic energy of repulsion between the condensed counterions.
To calculate the free energy of repulsion,
it is convenient to express $F^{\alpha \alpha}_n$ in terms of the
free energy of a one component plasma ($OCP$),  $n$ $\alpha$-ions
on the surface of a sphere with a uniform 
$neutralizing$ background, $F^{OCP}_n$.  The free
energy of a spherical $OCP$ can be written as 
\begin{equation}
\label{3} 
F^{OCP}_n=F^{\alpha \alpha}_n -\frac{\alpha^2 n^2 q^2}{\epsilon a} + 
\frac{\alpha^2 n^2 q^2}{2\epsilon a}\;.
\end{equation}
Substituting Eq.~(\ref{3}) into Eq.~(\ref{2}) the electrostatic free
energy of a polyion-counterion complex becomes,
\begin{equation}
\label{4} 
E_n=\frac{(Z-\alpha n)^2 q^2}{2 \epsilon a}+F^{OCP}_n \;.
\end{equation}
In the strong coupling limit, corresponding to multivalent counterions
or solvents of low dielectric permittivity, the free energy of the 
$OCP$ is well approximated by the free energy of the low
temperature phase corresponding to a triangular Wigner crystal,
\begin{eqnarray}
\label{5}  
F_{n}^{OCP}=-M\frac{\alpha^2 q^2 n^{3/2} }{ 2 \epsilon a }\;. 
\end{eqnarray}
where $M$ is the Madelung constant. For weaker couplings,
the expression for the $F_{n}^{OCP}$ can be obtained from the
fits to the Monte Carlo data~\cite{GaChCh79}.  
For  concreteness we shall use  
$M=1.106$, the value appropriate for a 
planar Wigner crystal~\cite{Le02}.

The effective charge of a polyion-counterion complex, in units of $-q$ is
\begin{eqnarray}
\label{6}  
Z_{eff}=Z-\alpha n \;, 
\end{eqnarray}
The  optimum number of condensed counterions 
is determined from the minimization of the total electrostatic 
free energy.  We find~\cite{Sh99a,MeHoKr01,Le02}
\begin{eqnarray}
\label{7}  
Z_{eff}^*=-\frac{1+\sqrt{1+4 \gamma^2 Z}}{2 \gamma^2} 
\approx -\frac{\sqrt Z}{\gamma}\;, 
\end{eqnarray}
where
\begin{eqnarray}
\label{8}  
\gamma=\frac{4}{3 M \sqrt \alpha}\;. 
\end{eqnarray}

We see that the optimal charge of a polyion-counterion 
complex is of opposite
sign to the bare colloidal charge, i.e. the complex is overcharged.  
Inside the colloidal suspension containing multivalent counterions
or solvents of low dielectric permittivity
the electrophoretic mobility can, therefore, be  reversed.

Some care, however, must be taken in extrapolating the results
of this simple model to real systems.
While we have treated the counterions as condensed on top of 
the sphere, this is clearly not the case for 
real colloidal suspension.  
Instead the associated counterions form a layer
around a colloidal particle which can be some nanometers wide.
The presence of simple electrolyte also strongly affects the
net charge of the polyion-$\alpha$-ion complex.  Furthermore, the
complex formation is a kinetic phenomenon requiring
a counterion to overcome an energy barrier in order to join
the already overcharged complex.  

\section{The overcharging potential}

In the previous section we found that the minimum of the total
electrostatic free energy of a polyion-$\alpha$-ion complex corresponds
to an  overcharged state. However, for
a counterion to join  an already overcharged complex it must overcome
an energy barrier. The waiting time  
for a thermal fluctuation of sufficient strength necessary 
to drive a counterion
over an activation barrier scales exponentially with the
height of the  barrier.  There is, therefore,  a kinetic limitation
to the  degree of overcharging which can prevent a thermodynamically
optimum state from being reached on  experimental time scale.
To explore this
further we have to construct an effective interaction potential 
between a complex and a counterion separated by distance $r$. 

The work necessary to bring a counterion from infinity to join
a complex containing $n$ $\alpha$-ions is
\begin{eqnarray}
\label{9}  
W=\frac{dE_n}{d n}\;. 
\end{eqnarray}
We define the reduced  
electrostatic potential of a counterion on the surface of the complex
as $\varphi(a)=\beta W$, where  $\beta=1/k_B T$.  
Differentiating Eq.~(\ref{4}) we find
\begin{eqnarray}
\label{10}  
\varphi(a) 
=-\frac{(Z-\alpha n) \lambda_B \alpha }{a} 
- \frac{3 M \alpha^2 \sqrt n }{4 a}\;, 
\end{eqnarray}
where $\lambda_B=q^2/\epsilon k_B T $.  
The first term of Eq.~(\ref{10}) is the 
electrostatic energy of interaction between a uniform spherical charge and 
an $\alpha$-ion, while the  second term is due to electrostatic
correlations between the $\alpha$-ions. 
In the strong coupling limit
correlational contribution to the interaction potential decay exponentially
fast with the separation from the polyion 
surface~\cite{RoBl96,LaPiLeFe00,StLeAr02}.  
The characteristic length is set by the average separation 
between the condensed counterions.  More specifically we can
approximate the reduced interaction potential by
\begin{eqnarray}
\label{11}  
\varphi(r)
=-\frac{(Z-\alpha n) \lambda_B \alpha }{r} 
- \frac{3 M \alpha^2 \sqrt n }{4 a} e^{-(r-a)/\xi}\;. 
\end{eqnarray}
The decay of the correlational contribution  
is governed by the characteristic length $\xi$ which in the
strong coupling limit is well 
approximated by~\cite{RoBl96,LaPiLeFe00,StLeAr02},
\begin{eqnarray}
\label{12}  
\xi=\frac{1}{|\mathbf G|}\;, 
\end{eqnarray}
where $\mathbf G$  is the reciprocal lattice 
vector of a triangular Wigner crystal of condensed counterions. 
Due to strong coupling between the condensed counterions, 
Eq.~(\ref{12}) should remain a good approximation
even significantly above the crystallization temperature.
For a triangular Wigner crystal,
\begin{eqnarray}
\label{13}  
{|\mathbf G|}=\frac{4 \pi}{\sqrt 3 b}\;, 
\end{eqnarray}
where $b$ is the lattice spacing
\begin{eqnarray}
\label{14}  
b=\frac{1}{3^{1/4} \sqrt \sigma }\; 
\end{eqnarray}
and $\sigma=n/4\pi a^2$ is the surface density of condensed counterions.
Substituting Eqs.~(\ref{13}) and (\ref{14}) into Eq.~(\ref{12}),
the decay length is found to be
\begin{eqnarray}
\label{15}  
\xi=\frac{3^{1/4}}{2 \sqrt{ \pi}} \frac{a}{\sqrt n}\;. 
\end{eqnarray}
We are now in  possession of the electrostatic potential
which will allow us to study the kinetics of overcharging.

\section{Kinetics of overcharging}

For $n<Z/\alpha$,  the electrostatic potential between
a counterion and a complex is purely attractive favoring further counterion
condensation.  
Inside an electrolyte solution this tendency towards polyion-counterion
association is opposed by the loss of entropy resulting  from the 
confinement of condensed counterions near the  colloidal
surface. 
Here, however, we shall not be concerned with the role of entropy~\cite{Le02}. 

For $n>Z/\alpha$ the interaction potential has two minima,
one located at $r=a$ and the second at $n=\infty$. For $Z/\alpha<n<n^*$
the $r=a$ minimum is the dominant one, 
while for $Z>n^*$ the global minimum
changes to $r=\infty$. The value of $n^*$ corresponds to the number of
condensed counterions which minimize the electrostatic free energy of
the complex Eq.~(\ref{4}),
\begin{eqnarray}
\label{16}  
n^*=\frac{Z-Z_{eff}^*}{\alpha}\;. 
\end{eqnarray}
\begin{figure}  
\begin{center}
\includegraphics[width=8cm, angle=270]{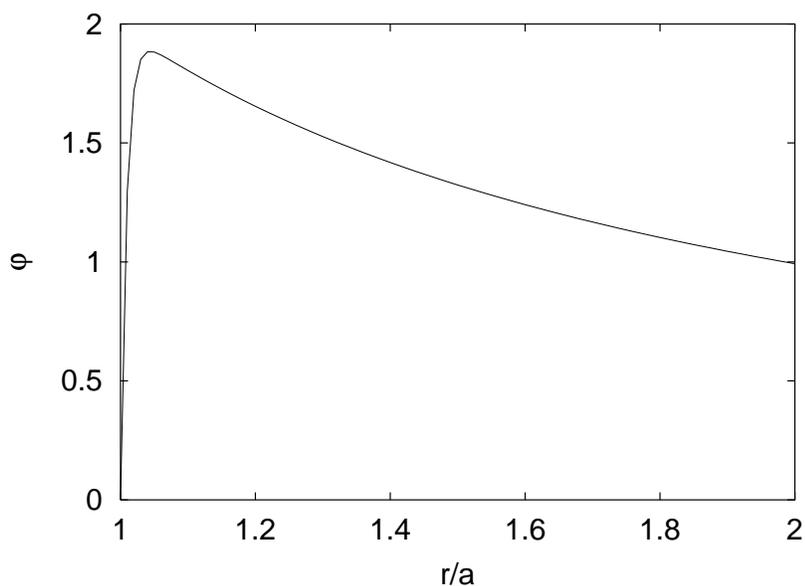}
\end{center}
\caption{The reduced interaction potential between a complex of $Z=4000$,
$a=1000$ \AA, $n=n^*$ condensed trivalent counterions,  and a trivalent
counterion located at distance $r$ from the center of colloid.}
\label{Fig1}
\end{figure}
In the case of trivalent counterions the energy
barrier that a counterion needs to overcome in order
to join a complex which already contains  $n^*$ condensed $\alpha$-ions 
is less than $2k_BT$, Fig 1.  Thus, for trivalent
counterions there is no kinetic hindrance to reaching the optimum
overcharged state. 

We next look at the height of the activation barrier as a function of
the counterion valence, Fig. 2. 
\begin{figure}  
\begin{center}
\includegraphics[width=8cm, angle=270]{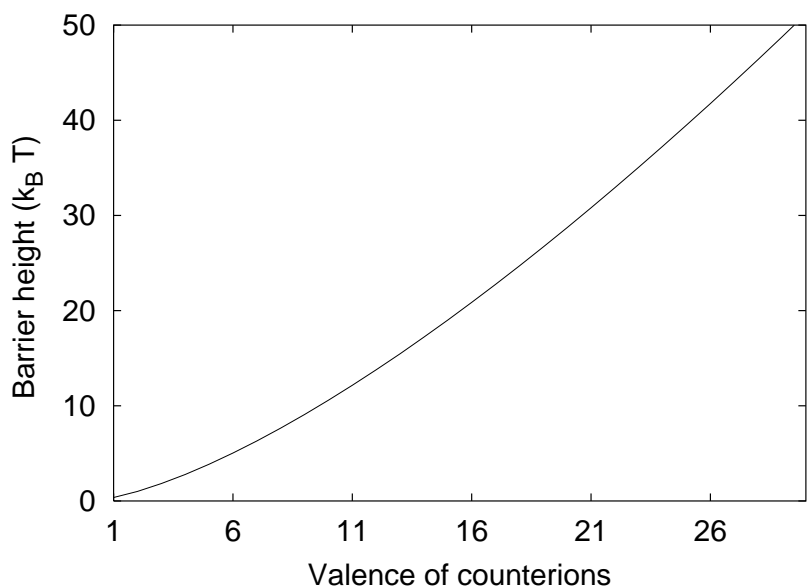}
\end{center}
\caption{The height of the activation barrier that an $\alpha$-ion
must overcome to join an optimally overcharged  complex
composed of a colloid with $Z=4000$, 
$a=1000$ \AA $\,$ and $n=n^*$ condensed $\alpha$-ions.}
\label{Fig2}
\end{figure}
It is clear that the height of the activation barrier
grows rapidly with the increased valence of the $\alpha$-ions. 
In particular we see that for $\alpha=10$ the activation barrier
is already some $10 k_B T$ which is probably the maximum
height that a counterion can overcome on a reasonable 
experimental time
scale. Thus, the process of overcharging by the $\alpha$-ions
with $\alpha>10$ will be kinetically controlled.
For example, from  Eq.~(\ref{7}) we see
that the optimal state of overcharging of a colloidal particle  
of $Z=4000$ and radius
$a=1000$ \AA $\,$ by micelles with $\alpha=25$ corresponds to 
$Z^*_{eff}=-271$.  In practice, though, the process of 
overcharging will come
to a stop when the barrier height reaches about $10k_B T$,
implying that the complex will stop growing when  
the net charge is only
$Z_{eff}=-70$.  

\section{Conclusion} 

In this contribution we have explored the kinetic limitation
to overcharging. We  find that kinetics does not play
an important role for overcharging by simple multivalent counterions,
so that the state of optimal overcharging, Eq.~(\ref{7}),  
is accessible within an 
experimental time scales. 
On the other hand, we find that the activation barrier grows
rapidly with the valence of counterions, suggesting that  
the extent of overcharging by
micelle-like aggregates is largely kinetically controlled.  

The kinetic limitation to overcharging might also be important
for the formation of  the DNA-cationic lipid complexes. The problem
of a reliable and safe  mechanism for gene  
delivery is particularly pressing in view of the current
medical applications.  Strong electrostatic
repulsion between a DNA and a 
cellular membrane inhibits transfection of a naked DNA into
the cell.  A way to overcome this difficulty is through
the  formation of  overcharged
complexes between the DNA and the cationic 
liposomes~\cite{FeRi89,Fe97,Fr97,HoMuAn98,KuLeBa99}.  These lipoplexes
having a net positive charge  are attracted to the cellular membrane,
facilitating the genetic transfection.

Finally,  the presence of a simple electrolyte will have a strong influence
on the overcharging. It has been demonstrated that for sufficient
concentration of $\alpha$-ions, monovalent 
salt favors overcharging~\cite{NgGrSh00c,Le02}.   In fact in the presence 
of simple electrolyte the thermodynamic state of optimum
overcharging corresponds to the charge inversion  of as much as $100\%$. 
This should be contrasted with the result of
Eq.~(\ref{7}), which shows that in the absence of salt, the
effective charge of a complex scales as 
a square root of the
bare charge.  The presence of salt 
will also lower the height of the activation barrier reducing the
kinetic hindrance to overcharging.

\newpage

\end{document}